\documentclass{article}

\PassOptionsToPackage{numbers}{natbib}

\usepackage[preprint]{neurips_2024}
\usepackage{graphicx}

\usepackage[utf8]{inputenc} 
\usepackage[T1]{fontenc}    
\usepackage[backref=page,hyperfootnotes=false]{hyperref}    
\usepackage{url}            
\usepackage{booktabs}       
\usepackage{amsfonts}       
\usepackage{nicefrac}       
\usepackage{microtype}      
\usepackage{xcolor} 
\usepackage{amsmath}
\usepackage{array}
\usepackage{caption}
\title{Lattice Lingo: Effect of Textual Detail on Multimodal Learning for Property Prediction of Crystals}

\author{%
  Mrigi Munjal$^*$ \\
  Department of Materials Science and Engineering\\
  Massachusetts Institute of Technology\\
  Cambridge, MA 02139 \\
  \texttt{mrigi@mit.edu} \\
  \And
  Jaewan Lee$^*$ \\
  LG AI Research \\
  Seoul, South Korea \\
  \texttt{jaewan.lee@lgreserach.ai} \\
   \And
  Changyoung Park \\
  LG AI Research \\
  Seoul, South Korea \\
  \texttt{changyoung.park@lgreserach.ai} \\
   \And
  Sehui Han \\
  LG AI Research \\
  Seoul, South Korea \\
  \texttt{hansse.han@lgreserach.ai} \\
}

\begin{document}

\maketitle

\begin{abstract}
  Most prediction models for crystal properties employ a unimodal perspective, with graph-based representations, overlooking important non-local information that affects crystal properties. Some recent studies explore the impact of integrating graph and textual information on crystal property predictions to provide the model with this "missing" information by concatenation of embeddings. However, such studies do not evaluate which type of textual information is actually beneficial. We concatenate graph representations with text representations derived from textual descriptions with varying levels of detail. These descriptions, generated using the Robocrystallographer package, encompass global (e.g., space group, crystal type), local (e.g., bond lengths, coordination environment), and semiglobal (e.g., connectivity, arrangements) information about the structures. Our approach investigates how augmenting graph-based information with various levels of textual detail influences the performance for predictions for shear modulus and bulk modulus. We demonstrate that while graph representations can capture local structural information, incorporating semiglobal textual information enhances model performance the most. Global information  can support performance further in the presence of semiglobal information. Our findings suggest that the strategic inclusion of textual information can enhance property prediction, thereby advancing the design and discovery of advanced novel materials for battery electrodes, catalysts, etc.
\end{abstract}

\def\thefootnote{*}\footnotetext{These authors contributed equally to this work}\def\thefootnote{\arabic{footnote}}
\begin{figure}
    \centering
    \includegraphics[width=1\linewidth]{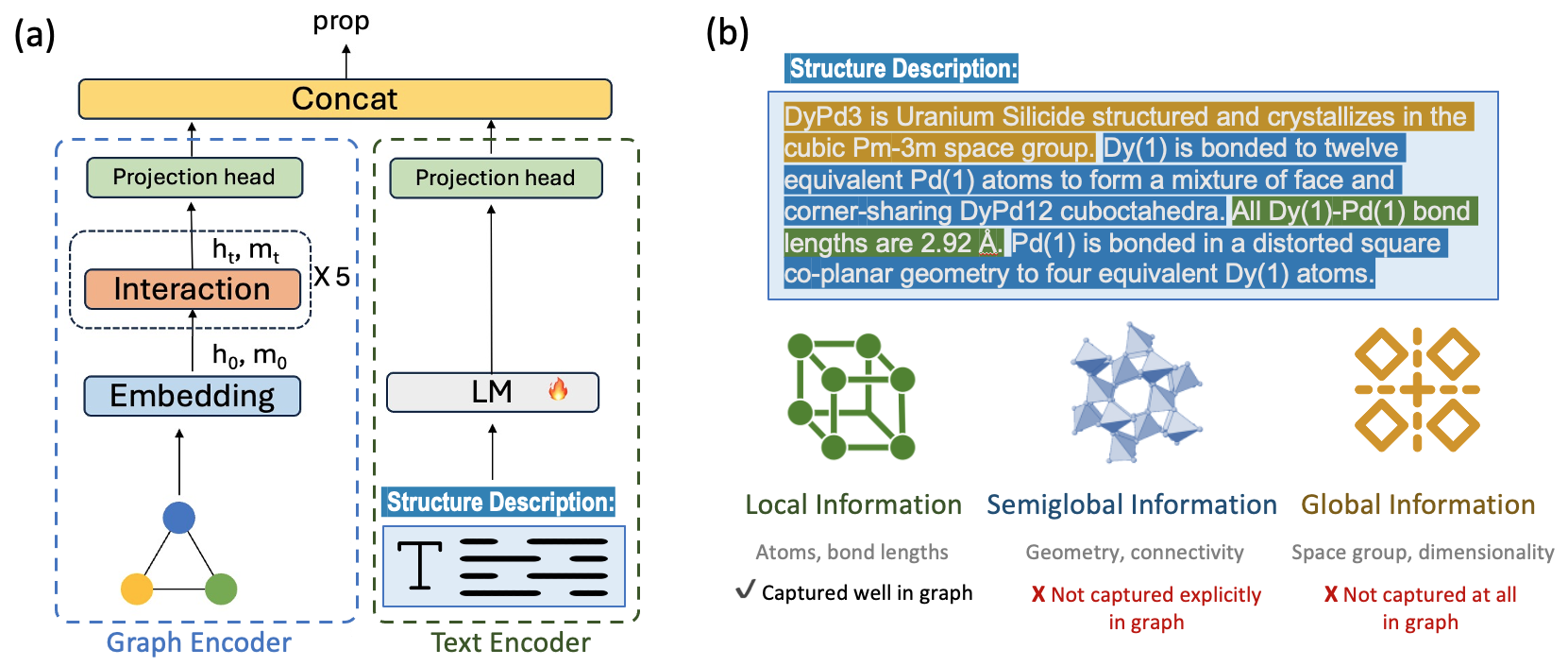}
    \caption{(a) The architecture of the multimodal model that combines graph and text representations through concatenation for property prediction. The left side showcases the coGN model with 5 interaction layers. `h' and `m' denote the node and edge embeddings, respectively. The right side showcases a Language Model (LM) to embed the structure descriptions. These embeddings are not frozen during training. These representations are concatenated and used for property prediction. 
    (b) Here is an example of a structure description from Robocrystallographer that has three levels of textual detail. In our experiments, we test how the model responds by varying this textual detail.}
    \label{fig:framework}
\end{figure}

\section{Introduction}

The discovery and design of novel materials rely on the ability to accurately predict the properties of crystal structures \cite{liu2017materials}. Recent approaches to property prediction with state-of-the-art performance have heavily relied on graph-based models that represent crystal structures as nodes (atoms) and edges (bonds or interactions), capturing the local atomic arrangements within these structures \cite{ruff2023connectivity, gasteiger2021gemnet}. These methods have been instrumental in understanding and predicting material properties by focusing on features like bond lengths, coordination environments, and local connectivity. Among the graph-based models, Connectivity Optimized Graph Networks (coGN) \cite{ruff2023connectivity} and Geometric Message Passing Neural Network (GemNet) \cite{gasteiger2021gemnet} have emerged as a powerful tools, excelling at capturing the connectivity of a crystal's structure through optimized representations. While these models excel in capturing short-range interactions, they often fall short in accounting for non-local information.

Non-local factors can significantly impact the physical and chemical properties of materials. For instance, symmetry information at the crystal level, which dictates how atoms are arranged in a repetitive pattern, can be crucial for understanding various material properties, including electronic band structure and mechanical behavior \cite{stoumpos2017structure, duffy2018single}. Similarly, the overall geometry and arrangement of atoms within a crystal can influence how materials interact with external fields or forces, impacting their mechanical and thermal properties \cite{callister2020materials, moulton2001molecules}. Another critical aspect is the layers or stacking information, particularly in polycrystalline and crystalline materials, that can affect properties like electrical conductivity, optical absorption, and more \cite{prince1995international}. 

Recognizing these limitations, some studies propose integrating graph-based structural representations with rich textual descriptions to enhance the prediction accuracy of crystal properties.  For instance, CrysMMNet is a multimodal framework that combines graph with textual descriptions to predict material properties \cite{das2023crysmmnet}.  Similarly, the Multimodal Learning for Materials (MultiMat) framework adopts a multimodal strategy by combining information from crystal encoders, density of states (DOS) encoders, and other property-specific encoders \cite{moro2024multimodallearningmaterials}. Although these approaches effectively merge graph representations with other modalities, they primarily focus on the integration of these modalities without delving into how information in each modality impacts the model's predictions.



 To address this, we systematically explore how the inclusion of different levels of textual detail affects the model's performance. This focus on the granularity of textual information allows us to better understand the contribution of the text modality and how it interacts with the graph modality to enhance property prediction. It also helps with identifying pathways for more efficient multimodal training by prioritizing the text that is most consequential and removing text that is not. Our approach particularly targets the prediction of properties like shear modulus and bulk modulus but can be used for other properties as well, like bandgap, formation energy, etc.

\section{Methods and Framework}
\label{gen_inst}

\subsection{Architecture}

The architecture we used integrates graph-based structural information with textual embeddings through concatenation similar to CrysMMNet \cite{das2023crysmmnet} as shown in Figure \ref{fig:framework} (a). This hybrid approach leverages the strengths of both modalities—graph-based models for capturing local atomic interactions and textual descriptions for non-local interactions.

For graph representations, we utilize the coGN model, which is adept at capturing local structural information within the crystal lattice \cite{ruff2023connectivity}. These representations are projected to obtain a 128-dimensional vector. Details of the coGN model are described in SI Section \ref{cogn_details}. To address the need for non-local information, we complement the coGN model with fine-tuned textual embeddings using MatSciBERT \cite{gupta2022matscibert}, a language model specifically trained on materials science literature. These embeddings are derived from textual descriptions generated by the Robocrystallographer package \cite{ganose2019robocrystallographer}, which encapsulate a wide range of information, including global characteristics (such as space group and crystal type), local details (such as bond lengths and coordination environments), and semiglobal features (such as connectivity and structural arrangements). These structured descriptions are converted into dense textual embeddings, which are also projected into a 128-dimensional vector. 

In the concatenation step, the vector from the graph representation is combined with the vector from the text representation. The concatenated vector, now enriched with comprehensive information from both modalities, is used to predict the material properties of interest. For the properties dataset, we used Materials Project data (v2023.09) \cite{jain2013commentary} and details of the dataset are described in SI Section \ref{tab:dataset}.

\subsection{Types of Textual Detail}
Textual descriptions can encompass a wide range of information, from local atomic environments to global symmetry and space group characteristics, each contributing uniquely to the prediction of material properties. As shown in Figure \ref{fig:framework} (b), the textual descriptions can include three types of information: local, semiglobal and global. These are described as following: 

\textbf{Local Information:}

Local information refers to the atomic-level details within a crystal structure, such as the atoms present and the specific bond lengths between atoms. Such information directly corresponds to the local atomic interactions that graph models are designed to encode.

\textbf{Semiglobal Information:}

Semiglobal information involves the geometry and connectivity within a crystal structure that extends beyond immediate atomic interactions but does not encompass the entire structure. This might include details about the arrangement of atoms in a broader region of the crystal, such as the connectivity between different atomic clusters or the orientation of specific polyhedra relative to each other.

For instance, textual descriptions might specify the overall geometry of a crystal's building blocks, such as "the DyPd\textsubscript{12} cuboctahedra are connected to form a distorted square co-planar geometry." This level of information is not typically explicitly captured explicitly by graph-based models. It plays a crucial role in determining properties that depend on the arrangement of multiple atoms or clusters, such as mechanical properties or electronic band structures. Incorporating semiglobal information through textual embeddings allows the model to account for these intermediate-scale features, which are essential for a more comprehensive understanding of material properties.

\textbf{Global Information:}

Global information refers to the overarching structural characteristics of a material, such as its space group, dimensionality, and symmetry properties. These features define the overall organization of the crystal lattice and can influence properties like electrical conductivity, thermal stability, mechanical behavior, and optical behavior. This type of information is not captured at all by graph-based models. Textual descriptions of global information might include statements such as "DyPd\textsubscript{3} crystallizes in the cubic Pm-3m space group." Moreover, we also experiment with appending descriptions of the space group to the Robocrystallographer generated information. Here is an example of space group description: \textit{P4/mbm: 'This space group contains a 4-fold rotation axis, two perpendicular 2-fold axes, and a combination of mirror and glide planes.'}


\section{Results and Discussion}
\label{headings}

The results of our experiments clearly demonstrate the value of integrating different types of textual information with graph-based structural representations to improve the prediction accuracy of key material properties. We evaluated the model performance on two important mechanical properties: Shear Modulus (log(G\textsubscript{VRH})) and Bulk Modulus (log(K\textsubscript{VRH})). The performance was measured using Mean Absolute Error (MAE), with lower values indicating better predictive accuracy. Table \ref{tab:experiment_results} summarizes the MAEs from our experiments. For the presented results, we used the average of 3-5 random seeds and each result of random seeds and  R2 scores are described in SI Table \ref{tab:experiment_results_R2}.

\begin{table}[ht!]
    \centering
    \captionsetup{justification=centering}
    \caption{This table showcases the MAE results for Shear and Bulk Moduli. Here `Graph Only' is the coGN model which is treated as baseline. `Description' refers to Space Group Descriptions.}
    \label{tab:experiment_results}
    \setlength{\tabcolsep}{2pt}
    {
        \begin{minipage}{0.49\textwidth}
            
            \centering
            \small
            \caption*{Shear Modulus (log(G\textsubscript{VRH}))}
            \begin{tabular}{lc}
                \toprule
                
                \textbf{Experiment} & \textbf{MAE} \\
                \hline
                \midrule
                \multicolumn{2}{l}{\textbf{Unimodal:}} \\
                 \midrule
                 Graph Only (Baseline) & 0.0883 \\
                 
                Full Text Only & 0.0825 \\
                \midrule
                \multicolumn{2}{l}{\textbf{Multimodal:}} \\
                \midrule
                Graph + Full Text & 0.0756 \\
                Graph + Local & 0.0872 \\
                Graph + Global & 0.0886 \\
                \textbf{Graph + Semiglobal} & \textbf{0.0749} \\
                Graph + Semiglobal + Global & 0.0759 \\
                Graph + Semiglobal + Global + Description & 0.0752 \\
                \bottomrule
            \end{tabular}
        \end{minipage}
        \hfill
        \begin{minipage}{0.49\textwidth}
            \centering
            \small
            \caption*{Bulk Modulus (log(K\textsubscript{VRH}))}
            \begin{tabular}{lc}
                \toprule
                \textbf{Experiment} & \textbf{MAE} \\
                \hline
                \midrule
                \multicolumn{2}{l}{\textbf{Unimodal:}} \\
                 \midrule
                 Graph Only (Baseline) & 0.0554 \\
                Full Text Only & 0.0455 \\
                \midrule
                \multicolumn{2}{l}{\textbf{Multimodal:}} \\
                \midrule
                Graph + Full Text & 0.0406 \\
                Graph + Local & 0.0516 \\
                Graph + Global & 0.0546 \\
                Graph + Semiglobal & 0.0399 \\
                \textbf{Graph + Semiglobal + Global} & \textbf{0.0385} \\
                Graph + Semiglobal + Global + Description & 0.0394 \\
                \bottomrule
            \end{tabular}
        \end{minipage}
        \hfill
        
    }
\end{table}

When using text representations of full text only, the model performance shows an improved MAE that graph representation only for both moduli, suggesting the importance of textual information. Combining both text and graph modalities results in an even better performance, underscoring the synergy. These findings suggest that textual descriptions can effectively complement graph-based information, filling in information for each other.

After testing the concatenation of graph representations with three types of textual detail: local, global, and semiglobal, it is evident that semiglobal information results in the best MAE improvement for both the moduli over the baseline. It also outperforms the "Graph + Full Text" case for both moduli. In contrast, local and global information barely support the graph modality. It is noteworthy that including only the relevant text information (semiglobal) is better than using the full text. Discarding inconsequential information can result in saved training and inference costs by reducing computation requirements. This outcome highlights the importance of semiglobal features, such as geometry and connectivity, which are not captured explicitly by graph models but are crucial for accurately predicting mechanical properties like shear modulus and bulk modulus. 

For shear modulus, the best performing model is graph-representation concatenated with semiglobal information. For bulk modulus, the most significant improvement was observed when combining structural information with semiglobal and global textual information unlike shear modulus, for which the model performs better without the global information. However, it is noteworthy that adding in space group description improves the utilization of the global information for shear modulus.  

The success of models that included semiglobal and global textual information indicates that properties like shear and bulk moduli are influenced by broader structural characteristics that are not captured by local atomic interactions alone. Geometry, connectivity, and symmetry play a critical role in determining the mechanical and physical properties of materials.



\section{Conclusion}
\label{others}

 Overall, our study highlights the significant advantages of integrating key textual information with graph-based models for predicting material properties. By systematically varying the granularity of textual descriptions—ranging from local atomic interactions to broader global structural characteristics—researchers can better understand how these different types of information contribute to the model's predictive performance for various material properties. The use of semiglobal and global information (in the presence of semiglobal information) has been shown to be crucial for capturing the complex dependencies that define properties in bulk materials. The use of this information over full text is also beneficial for efficient training by reducing the amount of textual information that needs to be processed.

By leveraging finetuned MatSciBERT embeddings, we have demonstrated how this approach can lead to improved predictions.   However, the smaller context window of BERT models can sometimes result in the truncation of textual information for some complex structures. Using LLMs like Mistral \cite{jiang2023mistral} and LLama models \cite{touvron2023llama} may improve performance but require significantly more resources for finetuning. Additionally, our current experiments are limited to shear and bulk modulus, but our methodology can be expanded to include other properties like bandgap and formation energy in the future.  We plan to address these avenues in our future work.


\begin{ack}
The authors thank the support from LG AI Research’s Global Internship Program. The authors also thank Prof. Elsa Olivetti, Dr. Kevin Huang, and Siddharth Nayak from MIT for their guidance on model development. 

\end{ack}

\small

\bibliographystyle{unsrt}
\bibliography{refs}

\medskip

\newpage
\appendix

\section{Appendix / supplemental material}

\normalsize

\subsection{Model Performance Details}

\begin{table}[h!]
    \centering
    \captionsetup{justification=centering}
    \caption{R2 results for Shear Modulus and Bulk Modulus}
    \label{tab:experiment_results_R2}
    \setlength{\tabcolsep}{2pt}
    {
        \begin{minipage}{0.49\textwidth}
            
            \centering
            \small
            \caption*{Shear Modulus (log(G\textsubscript{VRH}))}
            \begin{tabular}{cc}
                \toprule
                Experiment & R2 \\
                \hline
                 Graph Only (Baseline) & 0.884 \\
                Full Text Only & 0.895 \\
                Graph + Full Text & 0.907 \\
                Graph + Local & 0.882 \\
                Graph + Global & 0.875 \\
                \textbf{Graph + Semiglobal} & \textbf{0.908} \\
                Graph + Semiglobal + Global & 0.906 \\
                Graph + Semiglobal + Global + Description & 0.903 \\
                \bottomrule
            \end{tabular}
        \end{minipage}
        \hfill
        \begin{minipage}{0.5\textwidth}
            \centering
            \small
            \caption*{Bulk Modulus (log(K\textsubscript{VRH}))}
            \begin{tabular}{cc}
                \toprule
                Experiment & R2 \\
                \hline
                 Graph Only (Baseline) & 0.932 \\
                Full Text Only & 0.949 \\
                Graph + Full Text & 0.960 \\
                Graph + Local & 0.917 \\
                Graph + Global & 0.919 \\
                Graph + Semiglobal & 0.960 \\
                \textbf{Graph + Semiglobal + Global} & \textbf{0.962} \\
                \textbf{Graph + Semiglobal + Global + Description} & \textbf{0.962} \\
                \bottomrule
            \end{tabular}
        \end{minipage}
        \hfill
        
    }
\end{table}

\subsection{coGN Structure Encoder Details} \label{cogn_details}
The model starts with an Embedding layer that encodes the initial atomic features and connectivity information. This embedding serves as the starting point for the subsequent Interaction layers, which are designed to iteratively refine the representation of the material by passing messages between the nodes (atoms) in the graph. This interaction process is repeated five times, allowing the model to capture increasingly complex interactions within the material. Following the interaction steps, the refined graph representation is passed through a Projection Head that projects the learned features into a 128-dimensional vector.

\subsection{Limitations and Future Work}
Our study demonstrates the significant benefits of identifying key textual information that should be integrated with graph-based models for the prediction of material properties. This has opened up many opportunities that we could take on as future work. We could include other additional textual data, like detailed insights from X-ray diffraction (XRD) analyses or process-specific descriptions, that could further enhance the predictive capabilities of the model. The challenge moving forward is to systematically identify sources and incorporate these additional sources of textual information to provide a more comprehensive representation of material properties. ChemNLP \cite{choudhary2023chemnlp} as a tool may be able to provide part of this information. Although the model showed strong performance in predicting shear modulus and bulk modulus, its applicability to other material properties remains to be tested. 

 In addition to broadening the scope of textual information, optimizing the techniques used to embed this information will be critical. Exploring newer and more sophisticated language models, such as the Mistral 7B \cite{jiang2023mistral}, as well as the latest iterations of the Llama models \cite{touvron2023llama}, could lead to more accurate and contextually relevant embeddings. 

\subsection{Details of dataset}

\begin{table}[h!]
    \centering
    \captionsetup{justification=centering}
    \caption{Statistics of dataset }
    \label{tab:dataset}
    \setlength{\tabcolsep}{2pt}
    \begin{tabular}{|c|c|c|c|c|}
        \hline
        & \multicolumn{2}{c|}{\textbf{Shear Modulus log(G\textsubscript{VRH})}} & \multicolumn{2}{c|}{\textbf{Bulk Modulus log(K\textsubscript{VRH})}} \\ \hline
        \textbf{Data type} & \textbf{Train} & \textbf{Val/Test} & \textbf{Train} & \textbf{Val/Test} \\ \hline
        \textbf{Dataset size} & 7,513 & 1,910 & 7,513 & 1,910 \\ \hline
        \textbf{\# of nodes mean(std)} & 3.67(2.03) & 3.58(2.39) & 3.67(2.03) & 3.58(2.39) \\ \hline
        \textbf{\# of tokens mean(std)} & 223.99(120.76) & 220.97(117.84) & 223.99(120.76) & 220.97(117.84) \\ \hline
        \textbf{Text existence rate(\#)} & 100\%(7,513) & 100\%(1,910) & 100\%(7,513) & 100\%(1,910) \\ \hline
        \textbf{Y mean(std)} & 1.54(0.39) & 1.55(0.38) & 1.87(0.38) & 1.88(0.37) \\ \hline
    \end{tabular}
\end{table}

Table \ref{tab:dataset} summarizes the key statistics of the dataset for Shear Modulus and Bulk Modulus, partitioned into train and val/test sets. The number of nodes and tokens indicate the number of nodes in structure graph and text tokens in the text. There are cases where the Robocrystallographer is unable to generate text descriptions for certain materials. However, for the Shear Modulus and Bulk Modulus data, it was able to generate text for all materails. The means and standard deviactions of the properties are recorded in the Y mean(std) row.


\end{document}